# Possible Evolution of Supermassive Black Holes from FRI quasars

Matthew I. Kim[1,3], Damian J. Christian[1], David Garofalo[2], Jaclyn D'Avanzo[2]

1. Department of Physics & Astronomy, California State University, Northridge, CA 91330, USA
2. Department of Physics, Kennesaw State University, Marietta GA 30060, USA

Abstract

We explore the question of the rapid buildup of black hole mass in the early universe employing a growing black hole mass-based determination of both jet and disk powers predicted in recent theoretical work on black hole accretion and jet formation. Despite simplified, even artificial assumptions about accretion and mergers, we identify an interesting low probability channel for the growth of one billion solar mass black holes within hundreds of millions of years of the Big Bang without appealing to super Eddington accretion. This result is made more compelling by the recognition of a connection between this channel and an end product involving active galaxies with FRI radio morphology but weaker jet powers in mildly sub-Eddington accretion regimes. While FRI quasars have already been shown to occupy a small region of the available parameter space for black hole feedback in the paradigm, we further suggest that the observational dearth of FRI quasars is also related to their connection to the most massive black hole growth due to both these FRIs high redshifts and relative weakness. Our results also allow us to construct the AGN luminosity function at high redshift, that agree with recent studies. In short, we produce a connection between the unexplained paucity of a given family of active galactic nuclei and the rapid growth of supermassive black holes, two heretofore seemingly unrelated aspects of the physics of active galactic nuclei.

1. Introduction

The observation of powerful quasars requiring supermassive black holes in excess of a billion solar masses within several hundred million years of the Big Bang constitutes one of the most pressing unresolved issues in astrophysics (Begelman, Volonteri, & Rees, 2006; Genzel, et al., 2006; Jiang, et al., 2010; Valiante, Schneider, Salvadori, & Bianchi, 2011; Georgakakis, Rowan-Robinson, Babbedge, & Georgantopoulos, 2007; Melia, 2014; Volonteri M. , 2012; Greene J. E., 2012; Latif M. A., Schleicher, Schmid, & Niemeyer, 2013b). Since continued near-Eddington accretion with weak black hole feedback would be required, jumpstarting the process via seed objects such as massive population III stars (Nakamura & Umemura, 2001; Abel, Bryan, & Norman, 2002; Yoshida, Sokasian, Hernquist, & Springel, 2003; Gao, White, Jenkins, Frenk, & Springel, 2005; Pelupessy, Di Matteo, & Ciardi, 2007), massive stellar cluster collapse (Koushiappas, Bullock, & Dekel, 2004; Devecchi & Volonteri, 2009; Bellovary,

---

[3] Current institution: Jodrell Bank Centre for Astrophysics, University of Manchester, M13 9PL, UK. Email: wheelchair777@gmal.com

Volonteri, Governato, Shen, Quinn, & Wadsley, 2011; Lupi, Colpi, Devecchi, Galanti, & Volonteri, 2014), collapsed dark matter halos (Oh & Haiman, 2002; Mayer, 2010; Latif M. A., Schleicher, Schmidt, & Niemeyer, 2013a) as well as primordial supermassive black hole seeds (Carr et al. 2010; Johnson et al. 2012, 2013; Volonteri, Haardt & Madau 2003; Agarwal et al. 2012; Dijkstra et al. 2014; Booth & Schaye 2009; Argyres et al. 1998) is widespread. Unfortunately, the distribution of active galactic nuclei suggests sufficiently short duty cycles for active black holes that large seeds appear to also require a coupling to bursts of super-Eddington accretion in slim disks (Volonteri et al. 2015b; Abramowicz et al. 1988; Yu & Tremaine 2002; King 2003; Jiang, Stone & Davis 2014; Begelman, Volonteri & Rees 2006). Whereas continued accretion in thin disks with lower disk efficiency (i.e. retrograde disks) will satisfy the time constraints, the idea that mergers and/or secular processes may produce long-lasting, standard, low efficiency thin-disk feeding of a black hole in a restricted class of active galaxies, appears contrived and difficult to motivate, with little incentive beyond the simple imperative of forcing a solution.

Despite this recognition, we explore the buildup of supermassive black holes appealing to precisely such prolonged, standard, thin-disk accretion at only slightly sub-Eddington accretion rates, and mergers. Whereas we include increased accretion rates due to tilted disks during parts of the growth phase, our simple analytic exploration for the buildup of $10^9$ solar mass black holes in less than 800 million years from ordinary black hole seeds will appear contrived and unrealistic. We grant this. However - and here we find the primary reason for doing this - when interpreted within the context of the recent theoretical framework referred to as the gap paradigm, we find that a possible explanation emerges that not only addresses the black hole buildup issue, but also connects to the elusive Fanaroff-Riley Type I (hereafter, FRI) quasars, objects that are less common in our surveys of radio sources than we expect. We will show that at least a fraction of the FRI quasars are the kinds of objects that in the paradigm are naturally associated or compatible with the rapid buildup of black hole mass as a result of the fact that they emerge from a history of accretion involving weak feedback and higher redshifts, characteristics explaining their observational dearth. If a scenario that goes beyond our simplistic assumptions can be obtained, but that remains largely grounded in the basic ideas outlined in this paper, we are lead to an exciting prospect, namely that the central engines of the powerful, high-redshift quasars, must have experienced a past dominated by retrograde accreting black holes, which allowed them to grow both rapidly and quietly. But when these black holes lit up, they did so from recent ancestors in the form of relatively weak FRI quasars, high redshift, weaker jetted objects that have for good reason remained so far undetected.

This paper is presented as follows: In § 2, we produce a detailed analytic exploration of the feedback from accreting black holes in the recent gap paradigm in terms of mass-evolving black hole jet and disk powers, which will constrain the black hole buildup scenario. In § 3, we explore different tracks for the buildup of supermassive black holes involving mergers, tilted disks, and combinations of retrograde and prograde thin disk accretion (the full details of several of these scenarios are now in the appendices). While this hardly exhausts the physics relevant to the black hole buildup, we argue that it provides enough context to highlight the growth channel we are advocating in this paper. We will see that standard thin disk accretion, as is well known, is problematic in solving the high redshift black hole buildup problem, and identify a path for solving it via the aforementioned FRI quasar class of AGN. In § 4, we summarize and conclude.

## 2. Feedback from accreting black holes

The gap paradigm for black hole accretion and jet formation is a scale-invariant model for the jet-disk connection that distinguishes itself from the decades-old 'spin paradigm' by emphasizing the full retrograde-to-prograde range of accretion (Garofalo, Evans & Sambruna 2010). Of particular note in this model is an explanation to both the redshift distribution of radio galaxies and quasars, as well as the observation that radio loud quasars appear to be accreting as the analog to soft-state X-ray binaries which do not have powerful jets. For our purposes here, we focus on how the model prescribes tight constraints on the time evolution of AGN, with the most powerful jet-producing black holes in high retrograde spin configurations, where a range of black hole spin values for black holes threaded by magnetic fields on the order of $10^4$ Gauss are sufficient to explain the most powerful observed radio loud AGN. For the purpose of understanding rapid black hole buildup within the paradigm, our focus is on one aspect of the model connecting powerful Fanaroff-Riley Type II (FRII) quasar jets and their mode of accretion over time. If the FRII quasar jet is effective – an ability that depends on a combination of jet power and collimation – the model prescribes a heating of the galactic medium influencing both star formation as well as changing the state of accretion to hot mode, low angular momentum advection dominated accretion (ADAFs), from its original cold mode form. In the next section on the rapid buildup of black hole mass, we will show, that because ADAF accretion increases the timescale for mass buildup, succeeding in producing a billion solar mass black holes within hundreds of millions of years, by and large eliminates the ADAF mode of accretion from the buildup scenario, forcing us to understand the kinds of non-ADAF objects that are connected to FRII quasars in the paradigm. Therefore, we produce a precise determination of the jet power as the FRII quasar evolves in time by growing its black hole, allowing us to single out the objects that are both FRII quasars but whose jets do not appreciably change the accretion state, i.e. the objects whose radio mode feedback is inefficient. But the overall feedback from black holes has two components, one due to jets and the other to radiatively-driven disk winds, usually referred to as radio and quasar mode, respectively.

For the jet power, we explore the detailed time evolution of one accreting black hole using the following equation from Garofalo et al. (2010) by taking into account the mass that is accreted into the black hole.

$$L_{jet} = k\alpha\beta^2 m^2 j^2 B_d^2 \quad (1)$$

Where k is a spin-independent constant, m is the mass of the black hole, $B_d$ is the disk-threading magnetic field, and j is the normalized angular momentum,

$$\alpha = 2.5\left(\frac{3}{2} - j\right) \quad (2)$$

and
$$\beta = -\frac{3}{2}j^3 + 12j^2 - 10j + 7 - \frac{0.002}{(j-0.65)^2} + \frac{0.1}{(j+0.95)} + \frac{0.002}{(j-0.055)^2}. \quad (3)$$

Given an initial spin for the black hole, the jet power changes due to the fact that the black hole experiences changes in mass and spin values as angular momentum-carrying matter is accreted. The disk wind is also affected but in a way that is opposite that of the jet. As black hole mass increases, black hole spin evolves toward the greater prograde values and the innermost stable circular orbit (ISCO) moves inward toward the horizon, which affects both the disk luminosity and radiative wind power. The increase in black hole mass due to accretion of matter from the ISCO can be calculated as follows (e.g. Raine & Thomas 2005) and presented in Figure 1.

$$\Delta m = \int \frac{dm}{\left(1-\frac{2m}{3r_{ms}}\right)^{\frac{1}{2}}} \quad (4)$$

where $r_{ms}$ is a function of both mass and spin and represents the location of the ISCO in units where both the gravitational constant and the speed of light are equal to 1.

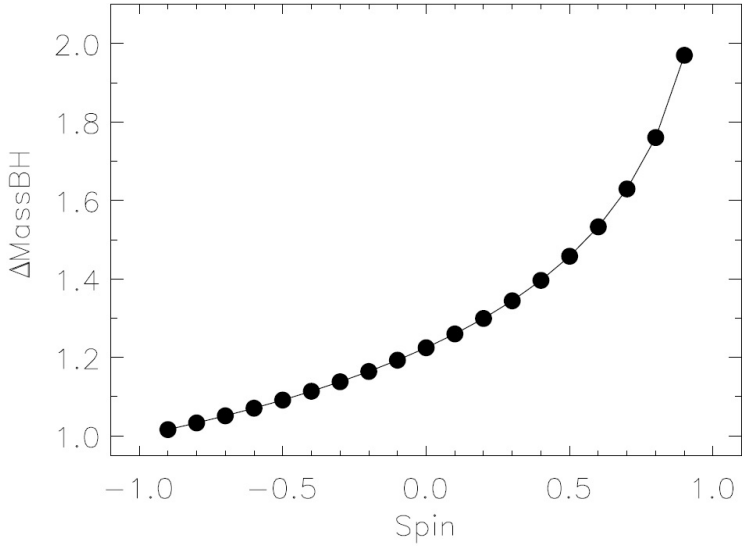

*Figure 1. Increase in black hole mass as a function of spin in terms of the original black hole mass.*

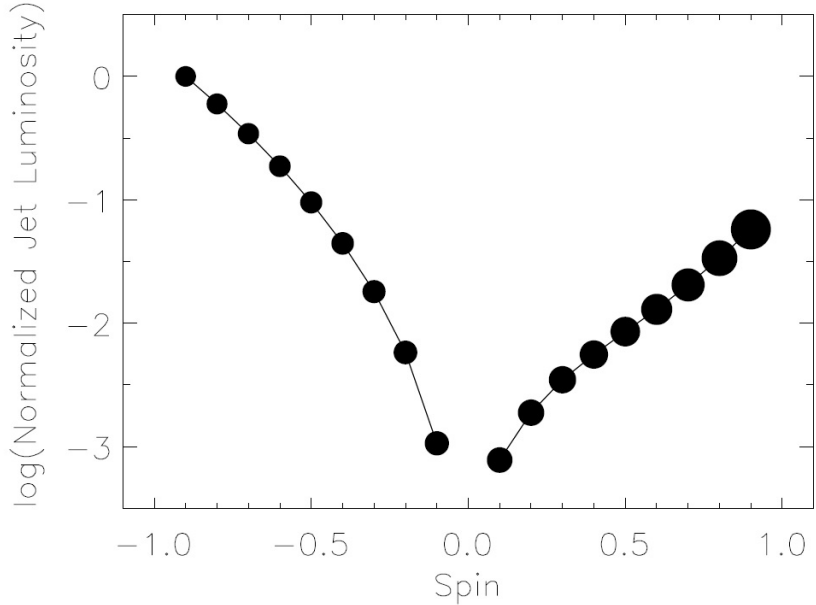

*Figure 2.* **Relative jet luminosity as a function of black hole spin. The size of the black circles captures the value of the black hole mass, i.e. the mass of the black hole at high prograde spin is about 3 times the mass of the original black hole at high retrograde spin.**

In Figure 1 we see the increase in mass acquired by the black hole as the spin changes by a finite fixed value equal to 0.1. In order to change the spin by the same fixed amount, the accreted mass must progressively increase in the prograde direction, i.e. more work is required by accretion to change the black hole's dimensionless angular momentum parameter. This simple feature of black hole accretion is the reason why prograde accreting black holes increase their mass by a factor of magnitude more than retrograde accreting black holes do, as accretion spins the black hole through the entire zero to maximum spin range. When we include this added black hole mass to the jet power expression, we find the values that appear in Figure 2. Whereas the black hole mass has changed by a factor approaching 3 as a result of the accreted matter from the disk as the spin changes from a value near -1 to a value near +1. The jet power in the prograde direction cannot reach the large values of jet powers produced in the retrograde spin range -1 to about -0.4, with the maximum jet power at the highest prograde value reaching only close to 0.1 times the jet power at highest retrograde spin.

In Figure 3 we show the normalized disk power as a function of black hole spin, i.e. the disk power in terms of the power at the highest retrograde spin value. Because the black hole mass increases according to Figure 1 as a result of the mass added via accretion, the disk power increases more, for a fixed spin difference, in the prograde regime. From standard thin-disk accretion theory, the disk power is

$$L = \frac{1}{2r} GM \frac{dM}{dt} \qquad (5)$$

where G is the gravitational constant, r is the radial location of the ISCO (which varies with black hole spin), M is the mass of the black hole (which increases with prograde spin), and $\frac{dM}{dt}$ is the accretion rate which we assume to be constant. In other words, the smaller the ISCO, the larger the overall power. And this translates into a greater radiative wind at a given radial location in the disk the higher the prograde spin value for a given black hole mass. When the accreted mass is taken into account during the time evolution, the increase in disk power in the prograde direction is even larger and captured in Figure 3, again with the size of the black dots representing the relative size of the black hole.

Figures 2 and 3 show how the entire retrograde regime is characterized by jet and disk powers behaving opposite to one another with respect to black hole spin, with the former decreasing and the latter increasing with increase in prograde spin. And the degree to which they do this from the perspective of the gap paradigm is calculated precisely here for one evolving accreting black hole. In the next section on the rapid buildup of black hole mass in the paradigm, we will be forced to consider multiple shots of retrograde accretion at near-Eddington rates which in turn will force us to come back to our jet and disk powers versus spin in an attempt to identify a class of AGN of the FRII morphology associated with weak radio mode feedback but increasingly greater quasar mode feedback. The end product of our rapidly growing black holes is a near-zero spin black hole surrounded by a slightly sub-Eddington accreting thin disk whose subsequent time evolution forces it into a prograde accreting regime, implying at least a brief life as a cold mode accreting FRI jetted AGN.

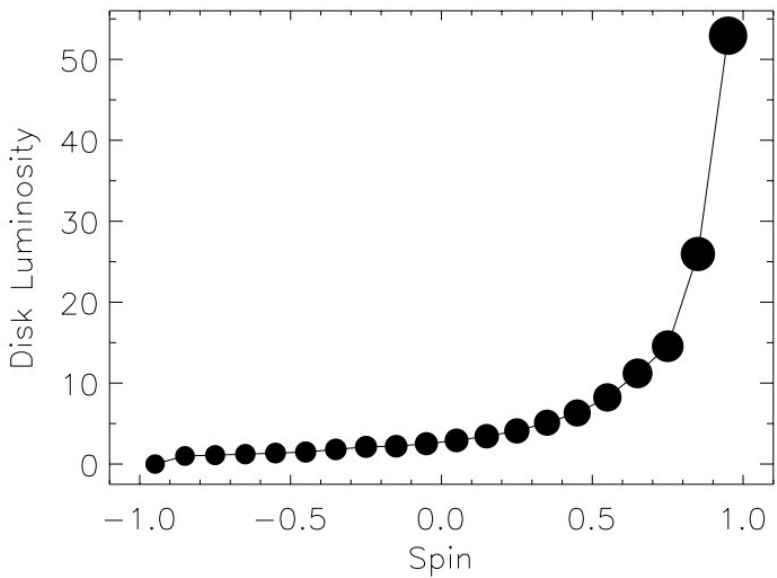

*Figure 3: Normalized disk power as a function of black hole spin.*

### 3. Supermassive black hole buildup

Because the overwhelming majority of supermassive black holes larger than a billion solar masses are associated with quasars at redshifts lower than 7, the small high redshift subgroup requires a plausible yet low probability explanation. Our goal in this section is to identify a scenario within the gap paradigm that satisfies these requirements. Whereas we will argue that the low probability aspect of our explanation emerges naturally, its plausibility is less obvious and much of our work will be to motivate it. The tools at our disposal will be standard, thin-disk accretion, in both prograde and retrograde configurations, mergers that double the black hole mass, and tilted disks for the initial post-merger accretion phase. We will explore simple scenarios taking into account self-gravity and disk-breakup, including estimates of timescales during and in-between the various physical processes. And, we will explore these paths in the context of the constraints imposed by the AGN feedback outlined in section 2. We are well aware of not doing justice to the entirety of physical processes thought to influence the growth of black holes and that even within the confines of the processes that we do employ, the scenarios that we explore are contrived. Instead, our goal is to constrain the space of possibilities to such an extent that a low probability, yet reasonable combination of physical processes, begins to emerge. Again, it is important to recognize that ours is not simply another proposal for solving the rapid buildup problem, but, instead, connects to unresolved AGN processes that seemingly have nothing to do with high redshift quasars, thereby producing (in our view) a richer and more predictive framework for understanding early massive black holes and their connection to the AGN family as a whole.

### 3.1. Mergers plus accretion

Without specifying the details of the seed black hole, we imagine a 100 solar mass black hole growing by accretion and mergers by 7 orders of magnitude via three tracks differing in the relative contribution of Eddington-limited retrograde and prograde accretion. The first track is governed by prograde accretion only. Here, we postulate mergers doubling the black hole mass with a post-merger zero spin black hole formed, surrounded by a standard thin disk accretion which is allowed to spin the black hole up to its maximum possible value, followed by yet another merger, again doubling the black hole mass, again with zero spin, again followed by the formation of a standard thin disk in a prograde configuration and so on until the buildup is complete. Whereas the mass doubles in each merger, the increase in mass due to accretion is about 2.45 times the original post-merger mass, while the timescale for the accretion spin-up at the Eddington limit is $10^8$ years. This timescale results from taking into account the luminosity of the disk whose efficiency changes with spin value. As the black hole spins up and the efficiency increases from about 6% of $\frac{dM}{dt}c^2$ to about 40% of $\frac{dM}{dt}c^2$, the luminosity increases which lowers the Eddington limit, forcing the system to accrete at lower rates, i.e. the Eddington-limited accretion decreases as the spin increases in the prograde direction. This process must continue until the mass has increased by the above mentioned factor of about 2.45. The total number of complete accretion events and mergers is therefore given analytically by

$$2.45^n 2^n M = 10^7 M \qquad (6)$$

which we solve to get

$$n = \frac{\ln(10^7)}{\ln(2.45)+\ln(2)} \approx 10 \qquad (7)$$

For understanding the contribution to the buildup of each process separately, let us at this stage only allow time to pass during the accretion phase, giving us a total buildup time of 10 x ($10^8$ yrs.) = $10^9$ yrs. = 1 billion years.

In short, despite the extremely contrived buildup scenario in which Eddington-limited accretion is followed instantaneously by a merger that always produces a zero spin black hole, we have violated the 800 million-year-buildup constraint. Before coming back to prograde accreting black holes with additional physics meant to make our scenario more realistic, let us explore the other two tracks in the same simplistic merger-plus-accretion context.

Whereas the prograde regime increases the original mass by a factor of about 2.45, a spin-down from maximum spin in a retrograde accretion configuration changes the mass by a factor of about 1.22 times the original mass (see section 2) with an Eddington-limited timescale of 8 x $10^6$ years. We can appreciate the different timescale compared to the prograde configuration in the following way. Because the innermost stable circular orbit is further away from the event horizon, the efficiency in retrograde configurations is less than in prograde ones, which in fact varies from 5-6% of $\frac{dM}{dt}c^2$. As a result of this, the accretion luminosity is lower for a given accretion rate compared to the prograde configuration and this ensures that the system can accrete at a greater rate for a given black hole mass compared to the prograde configuration, i.e. the equilibrium between radiative forces and gravity establishes itself at a higher net accretion rate. In addition to the greater accretion rate, the fact that innermost stable circular orbits are further away from the event horizon, ensures that the black hole acquires a greater angular momentum per unit mass per unit time compared to prograde configurations, which allows the black hole to go through the entire high spin to zero spin faster than in the prograde regime (i.e. 8 x $10^6$ years vs $10^8$ years). In short, Eddington-limited accretion in retrograde configurations evolves more rapidly compared to prograde configurations. Objections raised to prolonged retrograde accretion from the Bardeen-Peterson effect will be treated in the discussion. Here, we imagine that a maximally-spinning black hole is subject to a continuous accretion process that spins the black hole all the way down to zero and up again to high spin by the same accretion disk which now turns into the prograde configuration after the black hole has reached zero spin. Therefore, the number of mergers and accretion in this case is given by

$$2.45^n 1.22^n 2^n M = 10^7 M \qquad (8)$$

which can be solved analytically to get

$$n = \frac{\ln(10^7)}{\ln(2.45)+\ln(1.22)+\ln(2)} = 9 \qquad (9)$$

As we prescribed for the prograde-only track, let us only allow time to pass during the accretion phase, giving us a total buildup time of

9 x (8 x $10^6$ yrs. + $10^8$ yrs.) = 9.72 x $10^8$ years which also violates our 800 million year constraint.

Finally, let us consider mergers followed by retrograde-only accretion. The number of mergers is given by

$$1.22^n 2^n M = 10^7 M \qquad (10)$$

which can be solved analytically to get

$$n = \frac{\ln(10^7)}{\ln(1.22)+\ln(2)} = 18 \qquad (11)$$

Notice the large number compared to the prograde-only track. As mentioned previously, the amount of angular momentum per unit mass delivered to the black hole in the retrograde regime is larger than that in the prograde case. Hence, less total mass is accreted in retrograde regimes and more such accretion events are needed to build the black hole up to a billion solar masses compared to the prograde case. The timescale in this case is 18 x (8 x $10^6$ yrs.) = 1.44 x $10^8$ yrs. = 144 million years.

Within the confines of our extremely simplistic tracks, the retrograde-only path satisfies our time constraint.

There are several other physical effects that may influence the ability of a retrograde accretion scenario to build up a massive black hole which we explore in order to make the buildup more realistic. It is important to recognize, however, that the picture we wish to advertise for rapid black hole growth emerges from a foundation that is anchored in the simplest physical ideas that have now been well-established, namely accretion and mergers. While it goes without saying that we are not committed to the overly simplistic and fine-tuned picture that we construct in this paper, the take away message should be that an appeal to the simplest physical processes for black hole buildup remains an interesting option. Such effects that must be considered are self-gravity, tilted accretion disks and ADAFs. We now discuss the inclusion of these effects on our scenarios in the appendices. The bottom line is that, self-gravity and tilted disks produce negligible contributions to the mass buildup timescale, while ADAF are excluded from the buildup scenario. While tilted disks are not useful in lowering the timescale for black hole buildup appreciably, they will be essential in limiting the jet feedback from retrograde accreting black holes.

**3.2 FRI quasars as progenitors to high redshift SMBHs**

In order to constrain our attempt at identifying the non-ADAF objects that are compatible with our retrograde-dominated track, we begin by producing a quantitative analysis of the feedback from the black hole in the different tracks. We will find that the total feedback energy radiated from the accreting black hole in the retrograde-only track will be a less than unity fraction of the total feedback energy in the other tracks.

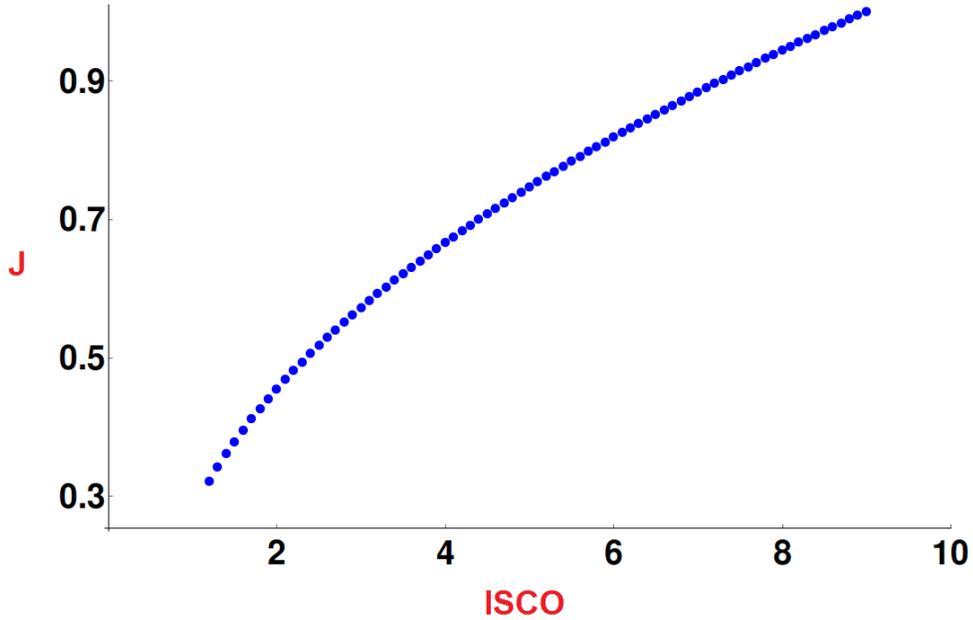

*Figure 4: Angular momentum per unit mass vs ISCO.*

In order to carry out this analysis, we appeal to Figures 2 and 3 which give us the jet and disk powers as a function of black hole spin and translate them into jet and disk power vs. time in order to integrate them and obtain the total jet and disk energies emitted during the black hole buildup. In the retrograde-only track, we found the need for 18 complete accretion events that spin the black hole down to zero from the initial high spin. The total energy emitted in jets is the area under the curve of a jet power vs time plot, which requires translating the horizontal axis of Figure 2 from spin to time. This is easily accomplished as we know the time associated with spin-down from high spin in the retrograde regime, namely less than $8 \times 10^6$ years if we allow tilted disks to deliver an appreciable amount of the angular momentum necessary to spin the black hole down to zero. However, the angular momentum per unit mass delivered to the black hole varies with ISCO values as shown in Figure 4, which means that changing the black hole spin at high retrograde values occurs faster for fixed accretion rate than at high prograde values. This means that translating a jet and disk power vs spin plot into a jet and disk power vs time plot requires stretching the horizontal space between values of spin progressively more as one moves toward the prograde regime. But the degree to which this stretching occurs depends on the rate of accretion. Our tracks begin with a tilted phase, which has the effect of increasing the accretion rate by a factor of 10, eventually leading into a phase where the disk accretes as a standard thin disk near the Eddington rate. We can understand the consequences of this as a rapid transition through the high retrograde spin values and a slower transition through the lower retrograde spin values. Quantitatively speaking, we obtain the values reported in Table 1 showing the ratio of total energy in the three tracks for both jets and disks.

| **tilted disks increase the accretion rate by 1 order of magnitude** | | |
|---|---|---|
| **Fractional jet energy** | **Fractional disk energy** | **Tracks being compared** |
| **0.20** | **0.015** | retrograde-only / retrograde+prograde |
| **0.22** | **0.014** | retrograde-only/prograde-only |
| **tilted disks increase the accretion rate by two orders of magnitude** | | |
| **Fractional jet energy** | **Fractional disk energy** | **Tracks being compared** |
| **0.02** | **0.017** | retrograde-only / retrograde+prograde |
| **0.02** | **0.019** | retrograde-only/prograde-only |

*Table 1: In the first column, first row, we compute the ratio of total energy produced in jets during the retrograde only track to the total energy in jets produced in the combined retrograde and prograde track. In the first column, second row, we compute the ratio of total jet energy produced during the retrograde track to the total jet energy produced in the prograde-only track. Note that both fractions are less than unity, indicating that the retrograde-only track produces less total energy from jets than in the other tracks despite the fact that it produces the most powerful jets. In the second column, we do the same for the total disk energy. Here we see an even more noticeable difference between the tracks with the retrograde-only track producing much less total output. The first two rows are obtained assuming that tilted disks enhance the accretion rate by only one order of magnitude. In the third and fourth rows, we compute the same quantities under the assumption that tilted disks increase the accretion rate by two orders of magnitude.*

The results of our feedback calculations reveal that our retrograde-dominated track produces a total energy in jets and disks that is about 20% and 1% of that in the other tracks, respectively. However, this 1% value for disks is obtained in the context of radiatively efficient, Eddington-limited, thin disk accretion. The analysis of our appendix forced us to conclude that our retrograde-dominated track will satisfy the 800 million-year time constraint even if the average accretion rate drops to 1/6 of the Eddington rate. If there were periods where the accretion rate drops to values that are even lower than that, the radiative efficiency of the disk would also drop, in turn decreasing the overall energy output further compared to the other tracks. For jets, on the other hand, we used tilted disks to deliver only half of the total available angular momentum to spin the black hole down. If we either increase the total angular momentum delivered during tilted phases and/or increase the accretion rate beyond the conservative 10% we adopted, total energy in jets will drop further compared to the other tracks. If we use an enhancement in the accretion rate during the tilted phase of three orders of magnitude and a total delivery of about 75% of the total angular momentum needed to spin the black hole down from high retrograde, we get an even smaller contribution in jet feedback while the disk feedback remains roughly stable as rows three and four of Table 1 show. We can understand this by recognizing that high retrograde configurations do not contribute significantly to the total disk energy so the radiative inefficiency of super-Eddington accretion during the short high retrograde phase has a relatively small contribution to disk energetics. The timescale for black hole buildup will not be lowered appreciably because 25% of the total angular momentum is still delivered in standard thin-disk form. In summary, we conclude counter-intuitively that the feedback from the rapidly growing track is less than in the slowly growing tracks. While we can appreciate the feedback result for disks in terms of disk efficiency being lower for retrograde black holes due to larger ISCO values, the jet feedback result is counter-intuitive because the jet efficiency in the paradigm is larger at high retrograde values. However, super-Eddington accretion ensures that powerful jets operate for short times. It is important to emphasize that our use of super-Eddington accretion does not contribute to resolving the issue of the time constraint per se; instead, it is instrumental in making the jet feedback weaker.

Our goal now is to identify within the paradigm the track that is compatible with the above feedback scenario that involves the FRI quasars. As we can see from Figures 2 and 5, the paradigm not only prescribes strongest jets in retrograde accreting black holes, but also an ability to alter the mode of accretion (Figure 5). Because we have concluded that rapid black hole mass buildup will not occur in the context of ADAF accretion, we need to look for further restrictions in the paradigm in an attempt to uncover the kinds of active galaxies that tend to remain close to Eddington-limited, thin-disk, accretion. Using Figure 5 we can see two extremes in the time evolution of FRII quasars, which are named FRII HERGs (High Excitation Radio Galaxy) in the figure for high excitation radio galaxies, indicating they are standard thin disks. The lower panel shows how the more powerful jet leads to a relatively rapid change in the accretion state from thin-disk to ADAF, with the spin value still in the retrograde regime when the accretion state has changed. On the top, we see the time evolution of a less powerful jet, which will not change the accretion state on rapid timescales so that the accretion state remains cold throughout the entire retrograde regime. In fact, the accretion state becomes an ADAF only when the system finds itself in the prograde regime. However, from our calculations in the previous section, we are not interested in systems that are close to ADAF accretion, since we were constrained to remain at an average accretion rate of about 1/6 the Eddington rate. Hence, in an average sense, we are interested in FRII quasars or FRII HERGs whose jets are even less effective in altering the mode of accretion than those depicted in Figure 5, such that even as they approach and enter the prograde regime, they are still characterized by cold, thin-disk, accretion. Note that in the prograde regime, the jets are of the FRI morphology.

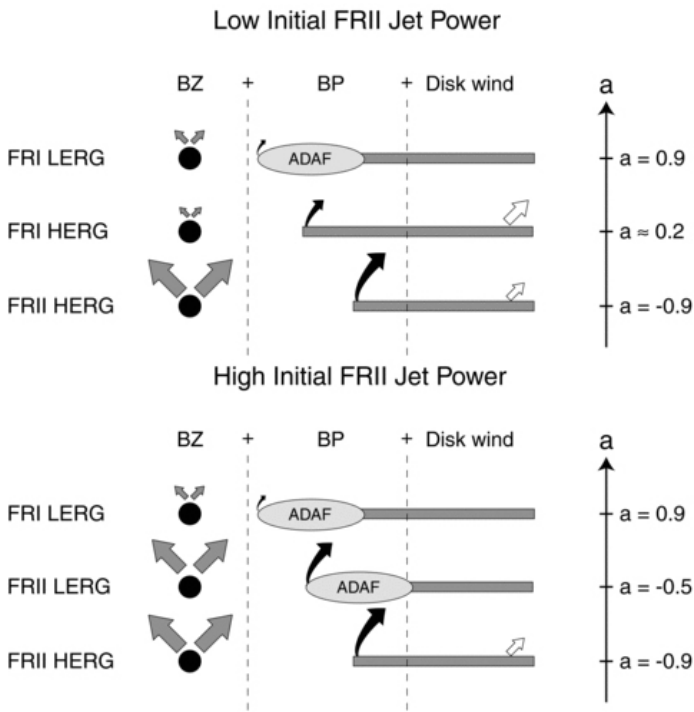

*Figure 5: Time evolution of an FRII quasar whose jet is either powerful enough to rapidly change the accretion state to ADAF (bottom panel) or less so (top panel), allowing the system to remain a thin disk throughout the retrograde spin down. Figure from Garofalo, Evans & Sambruna (2010). The arrows associated with 'BZ' refer to the power of the Blandford-Znajek black hole-driven jet, 'BP' to the power of the Blandford-Payne disk-driven jet, while the arrows associated with 'Disk Wind' indicate the strength of the radiatively driven disk outflow. HERG = high excitation radio galaxy (i.e. with thin disk accretion), LERG = low excitation radio galaxy (i.e. with ADAF accretion).*

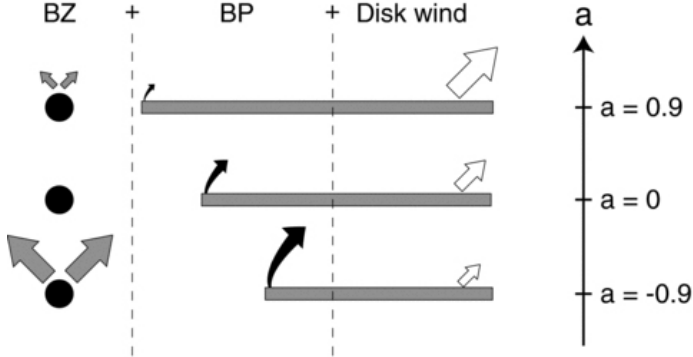

*Figure 6: An accretion disk that evolves from the high retrograde regime without a change in its accretion state. In this case, the initially high-spinning retrograde-accreting black hole evolves into the prograde-accreting regime while still in cold mode accretion. Since prograde accreting jets produce FRI morphologies in the paradigm and the disk is thin, such objects are referred to as FRI quasars.*

Therefore, we are interested in the class of retrograde jets that appear in Figure 6. Of course, we have constrained our buildup scenario with pure retrograde accretion, only interrupting the accretion process with a merger. Because such a scenario is unrealistic to an extreme, what we really should be contemplating is a retrograde-dominated scenario that occasionally also allows the prograde regime to enter the picture. The prograde regime, therefore, should not only not be excluded altogether, it is the end product of our retrograde-dominated scenario that contributes to our understanding of the evolution of our rapidly growing black hole. To the extent that accretion can at least to some extent continue in the low prograde spin regime, our end product, therefore, tends to be an accreting black hole with FRI jet morphology surrounded by cold mode accretion. These kinds of objects (FRI quasars) have been difficult to find (Blundell & Rawlings, 2001). In other words, whereas FRII quasars dominate the jetted population at redshift of about 2 and the FRI radio galaxies density increases at redshifts of about 1, the FRI quasars seem few and far between. What we are proposing here is that at least for a subclass of such objects – those associated with rapid black hole buildup - FRI quasars tend to emerge from accreting black holes whose jets are sufficiently weak in their feedback to fail to alter the accretion mode. Those are precisely the conditions of our retrograde-dominated track. The powerful FRII quasars that instead effectively alter the mode of accretion (Figure 4) would belong to a group of retrograde-accreting black holes that do not experience appreciable tilted disk phases.

With Figure 7 we provide a context for understanding our results within the larger picture of AGN evolution by constructing the high redshift part of the X-ray AGN luminosity function by connecting our results to those in Ranalli et al. (2015). What we have done here is to assume two 100 solar mass black holes beginning their buildup about 193 million years after the Big Bang at a redshift of 11, growing according to two of our tracks. The first is the mixed track involving a less fine tuned combination of retrograde and prograde accretion, while the other is the pure retrograde track that manages to grow to a billion solar masses within 800 million years. The discriminating factor between these two tracks involves the simple statistical likelihood that accretion events continue to form retrograde configurations in the aftermath of mergers and a relative probability is determined as a function of time. In other words, a given probability exists for the pure retrograde track to reach a 1000 solar mass black hole relative to the mixed track, an even lower probability exists for the pure retrograde track to reach a $10^4$ solar mass black hole and so on up until both tracks reach the billion solar mass value. Since the chance that every accretion event ends up in the retrograde direction gets smaller as the number of accretion events increases as the black hole builds up, the pure retrograde track becomes less likely over time. Because what we have determined is the relative probability between the pure retrograde track and the mixed track, including our results on the actual luminosity function forces us to make contact between the mixed track and the observed luminosity function. Once the mixed track is connected to the luminosity function, the pure retrograde track follows suit. Since the mixed track takes a greater time to reach a billion solar mass black hole, the hope is that it will be sufficiently long for it to enter a redshift range for which the luminosity function is known. In fact, the reason we choose the redshift of 11 to begin our buildup process is because based on that starting point, the mixed track reaches its destination at a redshift of 3, which is just within the range that allows us to make contact with the known luminosity function. In Figure 7, in fact, we show three tracks of the luminosity function taken from Ranalli et al. (2015) for luminosity classes greater than $10^{43}$ erg/s, greater than $10^{44}$ erg/s, and greater than $10^{45}$ erg/s in blue, green and black, respectively. And we indicate the members of these classes with increasingly larger circles indicating increasingly larger black holes. Clearly, the smallest luminosity class in blue has the smallest black holes while the largest luminosity class has the largest black holes in black. We make contact with the observed luminosity function by recognizing that our mixed track becomes a member of the three different luminosity classes as soon as its black hole is massive enough to produce the required luminosity to be a member of that class. We use the standard Eddington luminosity expression

$$L_{EDD} \approx 10^{38} \left(\frac{M}{M_\odot}\right)\frac{erg}{s} \qquad (12)$$

from which one can see that in order for our accreting black hole to enter into the first luminosity class, it must grow beyond $10^5$ solar masses from its initial 100 solar masses and so on until it reaches into the highest luminosity class. The red objects represent the mixed track with the smallest red disk indicating the location on the diagram where that object becomes a member of the smallest luminosity class as a result of the fact that its black hole has grown just beyond $10^5$ solar masses. And the second largest of the red points constitutes the location on the diagram where the mixed track object enters the second highest luminosity class, i.e. the green family. While the size of the red circles indicate the class in which they belong, the color is meant to emphasize where that object originates, i.e. how it has grown, in this case via the mixed track. Because it grows in a less fine-tuned way than the other track, it takes quite a bit beyond a billion years to reach a billion solar masses, and does so by a redshift of 3. We also wish to emphasize that it is because our framework does not allow rapid growth to occur in objects that have a broad spread in Eddington ratios - i.e. super-Eddington accretion does not contribute in any significant way to the black hole mass buildup - that we can use equation (12) to connect our mixed track to the observed luminosity function.

      The pure retrograde track, on the other hand, is indicated in yellow. It is important to note that the location of the largest yellow circle on the diagram is the most secure in terms of the theory since it is fixed in its location by the fact that it is connected to the largest red point, which in turn is itself securely fixed, related to it by a specific relative density determined by the relative probability mentioned above. In other words, once the red point is fixed in the highest luminosity class, the largest yellow point becomes fixed as well. The same is true for the other yellow points in relation to their corresponding red partners in size. However, the two smaller red points are located on the diagram in a way that is more uncertain which is by estimating an extrapolation from other members in their luminosity class. Unfortunately, our mixed track enters the smaller luminosity classes at too high a redshift to make contact with observations**.** Hence, the smaller red and yellow circles have larger uncertainties associated with them compared to the two larger ones. Note, finally, that our plot quantitatively captures the essence of our statement that the yellow track is a 'low probability channel'. In other words, the yellow track experiences the same physics that all our tracks exhibit (namely accretion of one type or other, and mergers), the only difference being the less likely combination of accretion types compared to the other tracks.

      As a final note, we comment on the recent history of how the Bardeen-Petterson effect (Bardeen & Petterson 1975) was thought to inhibit the formation of stable counter rotating disks, thereby invalidating our prolonged retrograde-dominated tracks. King et al. (2005) showed that the condition for the formation of a retrograde accretion disk is

$$\boldsymbol{\cos\theta < -\frac{J_d}{2J_h}} \qquad (13)$$

where θ is the angle between $J_d$ and $J_h$, the disk and black hole angular momenta, respectively. More recent work showed that under typical accretion disk conditions, not only are retrograde disks as likely to form as prograde ones in post mergers, they are stable (with the exception of self-gravity issues, the limitations of which we have already addressed) and can deliver the angular momentum to the black hole such that $\Sigma J_d = J_h$ as our tracks require (Nixon et al. 2011a,b; Nixon 2012). Although these simulations explored circumbinary accretion, these authors emphasize that the conditions for retrograde accretion around a single object or black hole are equivalent, a fact that opens up the interesting possibility of considering a coupling of our merger and retrograde accretion phases, thereby lowering the total time for the buildup of the black hole mass. In other words, we could have coupled the merger timescale with the time for retrograde spin-down by assuming that the retrograde disk provides a double function: 1) to merge the black holes by overcoming the final parsec problem (Nixon et al. 2011a) and 2) to spin the binary down to zero spin, producing a doubly massive black hole with no angular momentum. However, this would require a modification of the jet and disk powers in equations (1) and (5) and is therefore beyond our immediate scope. However, the idea suggests that the extra 10 million year average time added for each merger may be excessive or inappropriate.

In addition to this, there has also arisen the intriguing possibility of a black hole mass dependence on the formation of prolonged retrograde accretion, with black holes larger than about $10^8$ solar masses more easily spun down by accretion as a result of the fact that the disks around such massive black holes tend to have less total angular momentum (Dotti et al. 2013). This possibility offers an interesting resolution to the unanswered question of why radio loud AGN tend to host larger black holes compared to radio quiet AGN (Floyd et al. 2013). In other words, the more massive black holes (relative to their accretion disk mass) are more likely to remain in retrograde configurations if formed, which from the perspective of the gap paradigm allows them to produce the most powerful jets, evolving eventually into the giant FRI radio galaxies as described in Figure 4.

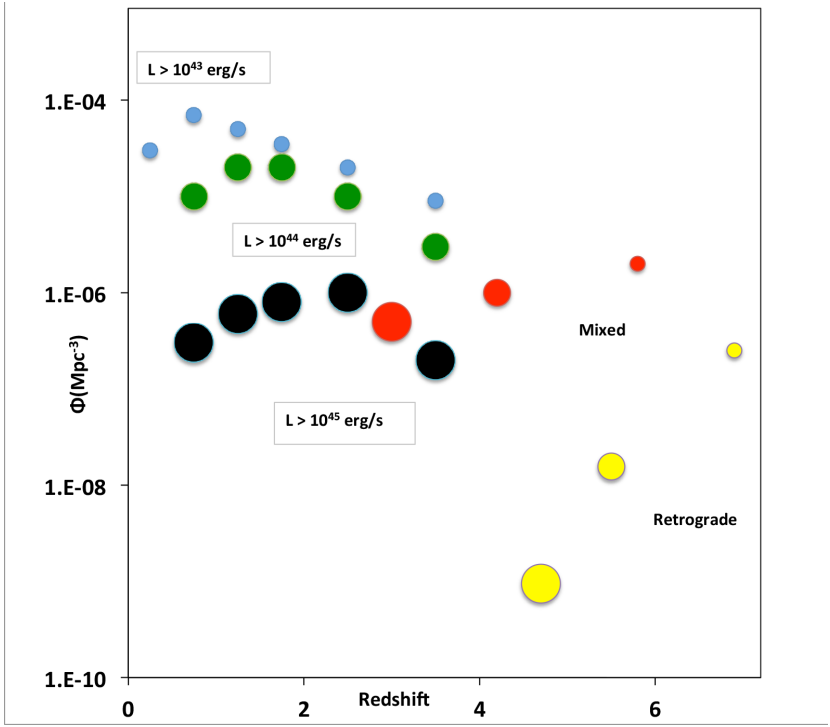

*Figure 7: The X-ray luminosity function for the mixed track (red) and the rapidly growing pure retrograde track (yellow). As the black hole grows rapidly via the pure retrograde track (yellow), the space density of such objects drops considerably compared to the mixed track (red). Whereas the largest yellow and red points are fixed by contact with observations, the other red and yellow points depend on an extrapolation to lower redshifts.*

In closing, we note that the bulk of the simulation work on misaligned disks has been carried out exclusively in the hydrodynamic regime. And general relativistic MHD simulations (GRMHD) suggest that the conditions for the Bardeen-Petterson effect are markedly different when electromagnetic torques are present (e.g. McKinney et al. 2013). However, the GRMHD work explored thick disks, so while not applying directly to our thin disk scenarios, it suggests that all the hydro-based studies should ultimately be carried out using a full MHD approach.

## 4. Summary and conclusion

In this paper, we have calculated the change in the relative jet and disk powers that result from following the increase in black hole mass due to accretion in order to explore within the paradigm a possible path for the buildup of massive black holes at early times in the Universe. Despite admittedly contrived accretion scenarios, we have argued for and motivated the idea that if thin-disk accretion matters, a mostly retrograde-dominated configuration is required to explain rapid black hole mass buildup. We have suggested that our simple calculations sufficiently highlight the nature of the ADAF-free path and found ourselves forced to explore FRII jets that are ineffective in their radio mode feedback due to short lifetimes, leading to an end product for the rapid buildup of black hole mass characterized by radiatively efficient thin-disk accretion around zero spinning black holes, systems that further accretion will spin up into a prograde-accreting thin-disk, which in the paradigm are objects belonging to the FRI quasar group.

The idea that is currently most subscribed to for the rapid buildup of black holes involves slim disks accreting at super-Eddington rates, sufficiently rapid to advect heat into the black hole, with the consequence of observationally lower radiative efficiency and quasar mode feedback. While this weak feedback nature of the rapidly growing black holes is also one we have appealed to, it appears in a different context, with super-Eddington accretion useful only in limiting jet feedback. But it is important to emphasize that our proposal does not spring from a simple search for weak-feedback AGN. Instead, it emerges from the need for two constraints that are independently satisfied by retrograde-dominated accretion, a time constraint and a feedback constraint. It is from this combination of weak feedback and retrograde-dominated accreting black holes that the FRI quasar class springs.

It is also important to emphasize that chaotic accretion – an idea that is commonly thought to operate in Seyfert galaxies to maintain low spin in black holes (King et al. 2006) - is not being advocated here, since such accretion is characterized by a combination of retrograde and prograde regimes. The powerful quasars we are modeling here involve an accretion history that is dominated by the retrograde regime but that is prone ultimately toward an end product involving continued accretion in the prograde regime, allowing the disk efficiency to reach maximum values, eventually leading to the most powerful quasar mode feedback. Whereas these rapidly growing black holes are relatively hidden by the combination of high redshift and weak feedback, efficient, very visible, quasar mode feedback is never too distant. In fact, if zero spinning black holes surrounded by radiatively efficient accretion is the end product of our growth track, such objects are capable of reaching the highest quasar mode feedback in about $10^8$ years at the Eddington rate. And this highest quasar mode feedback state must go through an FRI quasar phase as the accreting black hole crosses through the lower prograde spin regime. But once the prograde accreting black hole crosses the intermediate spin threshold value for jet suppression, the weaker radio mode feedback dies further, and the radio quiet nature of the quasar emerges fully. As previously explored, radio mode feedback (i.e. accreting black holes with non-negligible jets) in the prograde regime in radiatively efficient accretion, can only occur for intermediate spin values. At zero spin, the jet power is zero and at high prograde spin, the jet is suppressed (Neilsen & Lee, 2009; Ponti, Fender, Begelman, Dunn, Nesilsen, & Coriat, 2012a; Garofalo D. , 2013). Hence, the parameter space for FRI quasars is already small at any redshift.

In addition, our theoretical framework allows us to make contact with the observed X-ray luminosity function at redshift of above 3, thereby restricting the nature of the X-ray luminosity function and forcing a prediction from our model at highest redshift. It is also worth emphasizing that our model predicts that rapidly growing supermassive black holes are unlikely (i.e. the biggest yellow point (pure retrograde track) in Figure 7 corresponds to a $\Phi$ value of $10^{-9}$). While there are other unlikely fine-tuned growth tracks (such as prograde-only ones) that will not grow black holes rapidly, the ones that do are only part of the family of unlikely paths; i.e. there are no paths in the paradigm that are both more likely and that grow rapidly. By contrast, if super-Eddington-based explanations for large black holes at large redshifts are adopted, there need not be any obvious correlation between small values of Φ and large values of redshift. In our framework, instead, small Φ and large redshift are part and parcel of the same explanation. We also re-emphasize that we are not building a cosmological model for black hole growth in general, but one that is associated with rapid growth only.

In conclusion, this paper is an attempt to avoid a super-Eddington accretion-focused foundation for explaining the rapid black hole buildup in the early universe. In addition to the rapid black hole buildup, our framework allows us to also address the less advertised problem associated with the lack of observed FRI radio galaxies with cold mode accretion signatures, so-called FRI quasars. While we fully acknowledge that much of what we modeled is oversimplified, and that numerical simulations will ultimately have to explore this, our goal has been to produce what we find to be an intriguing connection between two aspects of the physics of extragalactic radio galaxies that have up to now never been explored in tandem. This is the latest in a series of recent work suggesting that retrograde accretion is useful in explaining a number of outstanding issues in high redshift AGN.

**Acknowledgements**

MK and DC thank the CSUN Department of Physics and Astronomy for support of this project. We thank C.W. Tsai for useful discussions on massive, high-redshift galaxies discovered by WISE. MK thanks to APJ and MNRAS referees for the depth to which they attempted to make sense of the ideas and we are all grateful for the improvements.

**Appendices**

**APPENDIX A. Mergers plus accretion plus self-gravity**

In this section, we add to our scenarios of section 3.1 the additional physics of self-gravity, which constitutes a limit on the amount of mass in the accretion disk compared to that of the black hole. If the mass of the accretion disk is greater than about $0.003 M_{BH}$, where $M_{BH}$ is the black hole mass, the self-gravity of the disk will tend to cause clumping, a break-up of the disk, and possibly star formation (Pringle, 1981; Gammie, 2001; Nixon, King, & Price, 2013). In a first attempt at dealing with this constraint, we imagine that our accretion disks come in chunks of about $0.003 M_{BH}$ (King et al. 2008). Let us begin again by exploring the pure prograde regime. Because the black hole mass will increase by 1.45 times the original mass, we need 1.45/0.003 = 483 accretion shots to get through one full prograde event. We ignore the fact that as the black hole increases in mass, the shots can increase in mass and still avoid being subject to self-gravity effects, and simply assume that all our shots have the same mass, equal to 0.003 the original black hole mass. In our attempt to account for all the timescales involves, we need to include a time in-between each shot. While ultimately negligible in the final account because we appeal to dynamical infall timescales, we add a nominal average timescale of $10^5$ years in-between each accretion shot where 0.003 of the original black hole mass is delivered, we have 483 x $10^5$ years = 4.83 x $10^7$ years as an additional time to complete one full prograde regime in addition to the pure accretion time which – because there are 10 complete accretion events - gives a time of 10 x 4.83 x $10^7$ yrs. = 4.83 x $10^8$ years for all 10 spin ups. Therefore, the total time to build a 1 billion solar mass black hole is now 1.483 billion years, still beyond our time limit.

For the mixed retrograde plus prograde accretion track we need 0.22/0.003 = 73 accretion shots to get through one full retrograde regime and an additional 1.45/0.003 = 483 accretion shots to get through one full prograde event. This gives a total of 73 + 483 = 556 shots. As we included in the pure prograde scenario, we assume an average time of 100 thousand years in-between each shot which adds a time of about 556 x 100 thousand years = 5.56 x $10^7$ years to complete one full retrograde plus prograde cycle. As we calculated in the previous section, the number of cycles in the retrograde plus prograde scenario is 9 so the time to complete 9 cycles is 9 x 5.56 x $10^7$ years = 5.004 x $10^8$ years. The time for pure accretion must be added to this, which we have already calculated above, giving us a total timescale to build a 1 billion solar mass black hole of 5.004 x $10^8$ years + 9.72 x $10^8$ years = 1.472 x $10^9$ years, again violating our time limit.

Finally, we explore the effect of self-gravity and the added time to the black hole mass buildup in the pure retrograde scenario where mergers lead to the formation of high-spinning merged black holes in retrograde accretion configurations. As we calculated in the mixed scenario, we need 73 accretion shots to complete one full retrograde event in order to ensure the disk stability. Using the same average time in-between shots of 100 thousand years, we have 73 x $10^5$ years = 7.3 x $10^6$ years added for the completion of one full retrograde event. We know that 18 such complete retrograde events are needed to build our black hole mass to 1 billion solar masses so we have 18 x 7.3 x $10^6$ years = 0.1314 billion years for all 18 spin downs and a total timescale of 0.1314 billion years plus the previously calculated 144 million years which leaves us at 2.754 x $10^8$ years. Hence, self-gravity does not break the time barrier in the purely retrograde regime. In the next section, we will continue our project of making the buildup more realistic by adding additional physics, that of tilted disks, which, while having the effect of reducing the timescales, do not have dominant effects.

**APPENDIX B. Mergers plus accretion plus self-gravity plus tilted disks**

We again begin with the purely prograde scenario whereby the merged product is a zero-spin black hole with twice the original mass surrounded by a thin accretion disk that spins the black hole up to maximum value. However, we now imagine that the initial accretion disk is thin but tilted and that it remains so for a time required to deliver half of all the accreting mass to the black hole in each shot. While GRMHD simulations of tilted but thick disks around rotating black holes report a modest increase in the rate of accretion (Fragile, et al. 2007), the increase in thin disks is more significant (Lodato & Pringle 2006) and in particular for misaligned disks that evolve toward retrograde configurations the accretion rate is enhanced by a factor > 100 (Nixon, King & Price 2012). We use a conservative average value for the increase in the accretion rate of a factor of 10. Therefore, the time for the accretion process is no longer $10^8$ years but 0.5 x $10^7$ years + 0.5 x $10^8$ years = 5.5 x $10^7$ years where the first term incorporates the factor of 10 decrease in time due to the factor of 10 increase in the rate at which mass is supplied to the black hole. Recalling that 10 complete accretion events are needed to build the black hole in the pure prograde scenario, the total time for accretion has changed in this case to $10 \times 5.5 \times 10^7$ years = $5.5 \times 10^8$ years. The in-between shots time is still $4.83 \times 10^8$ years which gives a total timescale to build the black hole of $4.83 \times 10^8$ years + $5.5 \times 10^8$ years = $10.33 \times 10^8$ years or 1.033 billion years. We are still hundreds of millions of years in excess of our time constraint. We can see that we can satisfy our time limit only if we assume tilted disks for the entire accretion process. But, again, our goal is to insist on an important contribution to the black hole buildup by standard thin-disk accretion.

Let us move on to the mixed retrograde plus prograde scenario in the context of tilted disks. Here, instead of a time of $8 \times 10^6$ years to spin the black hole down, we have a time of $4 \times 10^5$ years + $4 \times 10^6$ years = $4.4 \times 10^6$ years. In the prograde regime we take the time calculated above which is $5.5 \times 10^7$ years and obtain an accretion time of $4.4 \times 10^6$ years + $5.5 \times 10^7$ years = $5.94 \times 10^7$ years. Implicit in our numbers is the idea that each shot comes in tilted for exactly half of the total deliverable mass per shot. Given that the number of complete accretion events is 9, the accretion time for the retrograde plus prograde scenario is $9 \times 5.94 \times 10^7$ years = $5.346 \times 10^8$ years. To this we add the same in-between shots time of $5.004 \times 10^8$ years which gives a total timescale for the black hole buildup of $5.346 \times 10^8$ years + $5.004 \times 10^8$ years = 1.035 billion years. While this timescale is still in excess of our buildup constraint, it is now essentially the same as in the prograde scenario (i.e. only slight larger: 1.035 billion years vs. 1.033 billion years) due to the fact that there are twice as many opportunities to introduce tilted accretion during a complete cycle.

As usual, we finish the section with the pure retrograde scenario. Adding tilted accretion in the context of 18 complete accretion events gives an accretion time of $18 \times (4.4 \times 10^6 \text{ years}) = 7.92 \times 10^7$ years. The in-between time does not change from the previously calculated value of $1.314 \times 10^8$ years giving us a total timescale for the black hole buildup of $1.314 \times 10^8$ years + $7.92 \times 10^7$ years = $2.106 \times 10^8$ years which is still well within our 800 million year constraint. It is important to note that tilted disks have not appreciably lowered the timescale of formation of the massive black hole, which is a direct consequence of the fact that standard accretion operates for a time that is necessary to deliver an appreciable amount of the total angular momentum that is needed to drive the black hole spin to zero. In other words, either super Eddington accretion completely dominates the process or it has little weight in determining the order of magnitude timescale.

**APPENDIX C. Mergers plus accretion plus self gravity plus tilted disks plus merger time**

In this section, we include within our buildup tracks an estimate of the time associated with the black hole merger event. For our purposes, we are not interested in the details of the merging process, but simply wish to motivate an average order of magnitude that makes merger times neither negligible nor dominant. For this purpose we estimate an average timescale of about $10^7$ years. It is important to remember, however, that the early mergers involve isolated black hole binaries of hundreds of solar masses in protogalaxies whose environments are very different from those of major mergers involving half a billion solar mass black holes surrounded by gas rich environments. Our timescales, however, must ultimately be borne out in numerical simulations.

In the prograde-only track, we have 10 mergers, which amounts to an added time to the track of $10 \times 10^7$ years = $10^8$ years. We have a total time of 1.033 billion years plus the added $10^8$ years for all 10 mergers giving us a total of 1.133 billion years.

In the retrograde plus prograde track we have 9 mergers which adds a merger time of $9 \times 10^7$ years and a total time for the black hole buildup of 1.035 billion years + $9 \times 10^7$ years = 1.125 billion years.

In the retrograde-only track we have 18 mergers and a merger time of $18 \times 10^7$ years and a total timescale of $2.106 \times 10^8$ years + $18 \times 10^7$ years = $3.906 \times 10^8$ years, an admittedly contrived scenario that nonetheless satisfies our 800 million-year time constraint with 400 million years to spare. In the next and final section on the physics of mass buildup, we will make a connection between our black hole mass buildup scenarios and the prescription of the gap paradigm (and for that matter observations) that postulate the existence of a transition from cold, thin-disk, accretion, to hot, low angular momentum, ADAF accretion.

**APPENDIX D. Mergers plus accretion plus self gravity plus tilted disks plus merger time plus ADAFs**

In this section, we will constrain our buildup scenarios by invoking the aforementioned prescribed time evolution of the gap paradigm connecting ADAF accretion to FRII quasar jet phases (section 2). The basic feature of the model that we must consider is one which enforces a transition of the accretion phase from thin-disk to ADAF depending on whether or not the FRII quasar jet is sufficiently capable of heating the galactic gas in a way that influences the accretion phase (Garofalo, Evans, & Sambruna, 2010).

Because the pure retrograde track is alone in satisfying our time constraint for the black hole mass buildup that will be our focus here. We begin by allowing our retrograde accreting thin disks to evolve from standard, radiatively efficient, thin-disks, into hot, low accretion rate ADAFs. And because this transition is thought to occur at an accretion rate of about 0.01 of the Eddington accretion rate, this translates into a factor of more than 100 increase in the timescale for the delivery of a given accretion mass during the ADAF phase. As a point of comparison, let us look at what happens to the timescale for black hole mass buildup if we allow the mixed retrograde plus prograde scenario to experience a transition from thin-disk to ADAF for a non-negligible fraction of the accretion time. We imagine that the accretion rate has dropped to 0.005 and has turned into an ADAF. The timescale increase in the ADAF phase is now a factor of 200 larger than it was when we assumed it remained in a thin-disk configuration and the timescale has gone from less than 1.5 billion years to 190 billion years, making the buildup time larger than the age of the Universe by more than a factor of ten.

Instead of the $4 \times 10^5$ years + $4 \times 10^6$ years for the accretion timescale we used above for the retrograde-only track, we now imagine that the accretion disk has evolved into an ADAF phase following its tilted disk phase. This comes about in the paradigm as a result of powerful and collimated jets in the high retrograde spin regime. The spin has now evolved to about -0.5 and the disk has become an ADAF, which means we are dealing with low excitation radio galaxies with FRII morphology. Let us also assume the

5accretion rate has dropped to 0.005 of the Eddington accretion rate and the timescale has thus increased by a factor of 200. Considering all of the physics included in section 3.1-3.4, the ADAF phase has the most dominant effect, making the timescale to build the black hole up equal to $1.478 \times 10^{10}$ years which is slightly larger than the age of the observable universe. A latter half of the retrograde spin-down process, therefore, cannot be dominated by ADAFs. That's an inescapable conclusion. However, we can find the average accretion rate below the Eddington accretion rate that is compatible with the 800 million year time constraint. If we assume an average accretion rate during the second half of the retrograde spin down that is 1/6 the Eddington accretion rate, we find a total buildup time of 754 million years. But 1/6 of the Eddington accretion rate is almost 17 times the accretion rate that roughly serves as the boundary between thin, radiatively efficient, accretion, and ADAF accretion. The analysis in this section, therefore, forces us to the recognition that while lower average accretion rates are possible, the drop is not sufficient to allow ADAFs to enter the picture in any significant way.

**References**

Abel, T., Bryan, G., & Norman, M. (2002). *Science* , 295, 93-98.

Abramowiez, M. A., Czreny, B., Lasota, J. P., & Szuszkiewiez, E. (1988). *Astrophysical Journal* , 332, 646-658.

Agarwal, B., Khochfar, S., Johnson, J. L., Neistein, E., Dalla Vecchia, C., & Livio, M. (2012). *The Monthly Notices of the Royal Astronomical Society* , 425, 2854-2871.

Argyres, P. C., Dimopoulos, S., & March-Russell, J. (1998). *Physics Letters B* , 441, 96-104.

Bechtold, J., Siemginowska, A., Sheids, J., Czerny, B., Janiuk, A., Hamann, F., et al. (2003). *The Astrophyical Jurnal* , 588:119-127.

Begelman, M. C. (2002). *The Astrophysical Journal* , 568, L97-L100.

Begelman, M. C., Volonteri, M., & Rees, M. J. (2006). *The Monthly Notices of the Royal Astronomical Society* , 370,289-298.

Begelman, M. C., Volonteri, M., & Rees, M. J. (2006). *The Monthly Notices of the Royal Astronomical Society* , 370,289-298.

Bellovary, J., Volonteri, M., Governato, F., Shen, S., Quinn, T., & Wadsley, J. (2011). *The Astrophysical Journal* , 742, 13-19.

Bentz, M. C., Peterson, B. M., Pogge, R. W., & Vastergaard, M. (2009). Retrieved from Cornell University Library: http://arxiv.org/abs/0812.2284

Blandford, R. D., & Payne, D. G. (1982). *The Monthly Notices of the Royal Astronomical Society* , 199, 883-903.

Blandford, R. D., & Znajek, R. J. (1977). *The Monthly Notices of the Royal Astronomical Society* , 179, 433-456.

Blundell, K. M., & Rawlings, S. (2001). *The Astrophysical Journal* , 562, L5-L8.

Booth, C. M., & Schaye, J. (2009). *The Monthly Notices of the Royal Astronomical Society* , 398, 53-74.